\documentclass{sigchi}

\toappear{\small Permission to make digital or hard copies of all or part of this work for personal or classroom use is granted without fee provided that copies are not made or distributed for profit or commercial advantage and that copies bear this notice and the full citation on the first page. To copy otherwise, or republish, to post on servers or to redistribute to lists, requires prior specific permission and/or a fee. \newline   \emph{WebSci'14}, June 24 -- June 26, 2014, Bloomington, Indiana.\\ ACM 978-1-4503-1889-1.}
\usepackage{balance}  
\usepackage{graphics} 
\usepackage{times}    
\usepackage{url}      
\usepackage{CJKutf8}  

\makeatletter
\def\url@leostyle{%
  \@ifundefined{selectfont}{\def\UrlFont{\sf}}{\def\UrlFont{\small\bf\ttfamily}}}
\makeatother
\def\pprw{8.5in}
\def\pprh{11in}

\usepackage[pdftex]{hyperref}
\hypersetup{
pdftitle={WebScience Conference Proceedings Format},
pdfauthor={LaTeX},
pdfkeywords={WebScience, proceedings, archival format},
bookmarksnumbered,
pdfstartview={FitH},
colorlinks,
citecolor=black,
filecolor=black,
linkcolor=black,
urlcolor=black,
breaklinks=true,
}

\begin{document}

\title{Prediction of Foreign Box Office Revenues Based on Wikipedia Page Activity}

\numberofauthors{2}
\author{
  \alignauthor Brian de Silva\\
    \affaddr{HRL Laboratories, LLC.}\\
    \affaddr{3011 Malibu Canyon Road}\\
    \affaddr{Malibu, CA 90265}\\
    \email{bdesilva@hrl.com}
  \alignauthor Ryan Compton\\
    \affaddr{HRL Laboratories, LLC.}\\
    \affaddr{3011 Malibu Canyon Road}\\
    \affaddr{Malibu, CA 90265}\\
    \email{rfcompton@hrl.com}
}


\maketitle

\begin{abstract}
A number of attempts have been made at estimating the amount of box office revenue a film will generate during its opening weekend. One such attempt makes extensive use of the number of views a film's Wikipedia page has attracted as a predictor of box office success in the United States. In this paper we develop a similar method of approximating box office success. We test our method using 325 films from the United States and then apply it to films from four foreign markets: Japan (95 films), Australia (118 films), Germany (105 films), and the United Kingdom (141 films). We find the technique to have inconsistent performance in these nations. While it makes relatively accurate predictions for the United States and Australia, its predictions in the remaining markets are not accurate enough to be useful.
\end{abstract}




\section{Introduction}

The motion picture industry is a multi-billion dollar market.
Therefore many attempts have been made at predicting how much money films
will bring in during their respective opening weekends in theaters \cite{Huberman:twitter, mestyan:wikipedia, sharda:predicting}. There is also a body of research on the effects of various factors on box office revenue which assists those attempting to predict it \cite{basuroy:critical, liu:word, terry:determinants, zhang:improving}.

Huberman et al. found that they could use chatter on Twitter\footnote{\url{https://twitter.com/}} about movies prior to their release to accurately predict how well they would do in theaters \cite{Huberman:twitter}. More recently, Mesty{\'a}n et al. successfully used data gathered from Wikipedia\footnote{\url{http://www.wikipedia.org/}}, including page views, total edits, and number of editors, to predict the amount of money a film (released in the United States) would bring in during its opening weekend \cite{mestyan:wikipedia}.

It is a rather unexpected result that the number of views a film's Wikipedia page has is so highly correlated with the money earned by that film during its opening weekend.
A possible explanation for this phenomenon is as follows. In modern society potential moviegoers who are unsure which movie to see often turn to the internet to aid them in making their decision. 
It was recently suggested by Panaligan et al. that opening weekend box office sales for a film could be well approximated by the search volume for terms related to the movie. They were able to account for as much as 94\% of the variation in opening weekend box office performance using a combination of related search term volume, franchise status, and seasonality \cite{panaligan:google}. If we take this result into account along with the observation that upon searching for a film (using \textit{Google} or \textit{Bing}), that film's corresponding Wikipedia page is among the top search results, it is reasonable to hypothesize that the Wikipedia page acts as a proxy for searches relating to the film.

Our goals in this article are to recreate the results of Mesty{\'a}n et al. using a more recent set of films released in the United States and then to test whether or not the technique achieves comparable results when applied to films released in the United Kingdom, Australia, Japan, and Germany. First we discuss the methods we used to complete our study and then we present the results accompanied by a brief analysis.

\section{Methods}
\label{methods}
In order to carry out this study, we used a set of $325$ movies released in the United States in 2013, $141$ movies released in the United Kingdom in 2013 and 2014, $118$ movies released in Austrialia in 2013 and 2014, $95$ movies released in Japan in 2013, and $105$ movies released in Germany in 2013. We retrieved the box office data for the films released in each country from Box Office Mojo \footnote{\url{http://www.boxofficemojo.com/}}. From the same site we were also able to gather data on the number of screens each film was played on while in theaters (this data is often available prior to a film's release date).
Determining the web address of each film's Wikipedia page proved to be much more difficult in comparison. Our solution was to develop an automated procedure to accomplish this task (see the \textit{Finding Wikipedia URLs} Section).

Once we had the URLs of the Wikipedia pages corresponding to the films we were interested in, we used a website which records Wikipedia article traffic statistics\footnote{\url{http://stats.grok.se/}} to download the daily view counts of each page (see the \textit{Regression Model} Section). We then used a simple multivariate linear regression to attempt to predict each film's opening weekend box office revenue based on its Wikipedia page views and screen count. Finally, we analyzed the accuracy of the regression model using leave-one-out cross-validation (in the \textit{Cross-Validation} Section).

\subsection{Finding Wikipedia URLs}
\label{wiki urls}

Wikipedia URLs unambiguously identify films. Other sensors of public sentiment used for box office prediction, such as mentions of titles in Twitter text \cite{Huberman:twitter} or Google search queries \cite{panaligan:google}, are difficult to work with for titles based on common words (e.g. a tweet mentioning "Frozen" may not be referencing the film). Disambiguation becomes substantially more difficult when long lists of titles in multiple foreign languages are involved.

Wikipedia articles seldom contain information about opening weekend box office. In order to conduct our research we therefore need to construct a mapping between the opening weekend box office data available from Box Office Mojo and the associated language-specific Wikipedia URLs. Since Wikipedia URLs are not necessarily exact title names (e.g. \url{de.wikipedia.org/wiki/42_(Film)} ), this mapping must be constructed by hand (which is prohibitively time consuming) or through an automated fashion.

We check lists built from DBpedia\footnote{\url{http://dbpedia.org/}} against results obtained via the Google Custom Search API\footnote{\url{www.google.com/cse}} to resolve spelling differences between titles on Box Office Mojo and Wikipedia URLs. For each of the nations considered, we restrict the search API to to return links only from the national language's Wikipedia (i.e. \url{en.wikipedia.org}, \url{ja.wikipedia.org}, \url{de.wikipedia.org}) and query the search API for Box Office Mojo title names followed by the appropriate translation of the word ``film'' (i.e. ``film'' (en), ``film'' (de), or ``\begin{CJK}{UTF8}{min}映画\end{CJK}'' (ja) ).

Despite the addition of ``film'' to our searches, the highest ranked search results are not necessarily the Wikipedia pages for the queried films. 

To ensure that the links retrieved from our searches indeed refer to the Wikipedia URLs for each film, we use language-specific variants of DBpedia \cite{auer2007dbpedia}
 to build lists of URLs which are categorized as films and discard search results which do not appear in the lists.

For example, when entered into \url{http://ja.dbpedia.org/sparql}, the query below provides us with a list of 769 Wikipedia URLs which describe films (Category:\begin{CJK}{UTF8}{min}2012年の映画\end{CJK} ) or animated films (Category:\begin{CJK}{UTF8}{min}2013年のアニメ映画\end{CJK} ) which were released in Japan during 2012 or 2013:

\begin{CJK}{UTF8}{min}
\begin{verbatim}
PREFIX c: <http://ja.dbpedia.org/resource/
	Category:>
PREFIX dcterms: <http://purl.org/dc/terms/>
SELECT ?film WHERE {
    {?film dcterms:subject c:2012年の映画 .}
    UNION
    {?film dcterms:subject c:2013年の映画 .}
    UNION
    {?film dcterms:subject c:2013年のアニメ映
    画 .}
    UNION
    {?film dcterms:subject c:2012年のアニメ映
    画 .}
}
\end{verbatim}
\end{CJK}

These queries provide us with lists of Wikipedia URLs for: 769 Japanese films, 1026 German films, and 3079 English films.

Opening weekend box office data on Box Office Mojo is available for: 104 Japanese films, 166 German films, 219 Australian films and 225 UK films.

The technique provided here allows us to align Wikipedia URLs with box office data for: 73 Japanese films, 132 German films, 118 Australian films, and 141 UK films. Some alignments were also performed manually.

\subsection{Regression Model}
\label{regression}
As previously mentioned, we used a multivariate linear regression model\footnote{We used the linear regression (ordinary least squares) implementation provided by the \textit{Scikit-learn} Python package (\url{http://scikit-learn.org/stable/index.html})\cite{scikit-learn}.} of the following form to predict the revenue generated during each film during its opening weekend:
\begin{displaymath}
y_i = \alpha _1 x_{i,1} + \alpha _2 x_{i,2} (t) + \varepsilon _i ,
\end{displaymath}
where $y_i$, $x_{i,1}$, $x_{i,2}$, and $\varepsilon _i$ are the opening weekend box office revenue, screen count, number of Wikipedia page views\footnote{We chose to use the cumulative page views, i.e. the sum of the number of times the page was visited starting at some fixed number of days before release, up until day $t$.}, and the error in the prediction, for film $i$, respectively. $\alpha _1$ and $\alpha _2$ are the regression coefficients. Notice that $x_{i,2}$ is a function of time in the above equation. Since we have a range of dates for which we know how many views each film's Wikipedia page received, we have the ability to form a large number of regression models, one for each day leading up to the film's release. For example, we may wish to predict a movie's box office success using only data available one month before it premiers. Collecting the Wikipedia page hits over a long period of time allows us to discern how the accuracy of the regression model changes over time, as more information is introduced. One quantity we will use to measure accuracy is the R$^2$ Coefficient of Determination. We also found it prudent to use a second tool to evaluate our method's performance.

\subsection{Cross-Validation} 
\label{validation}
Due to the large amount of potential training data available on movie ticket sales and Wikipedia page views, we decided that leave-one-out cross-validation was an appropriate technique for assessing the performance of our model\footnote{We used the leave-one-out cross-validation implementation provided by the \textit{Scikit-learn} Python library (\url{http://scikit-learn.org/stable/index.html}).}. Given a set of films along with their associated box office revenues, Wikipedia page views, and screen counts, we successively remove one film from consideration, form a linear regression model using the remaining films, and then use the resulting model to predict the box office revenue for the film we removed\footnote{Note that leave-one-out cross-validation is an appropriate method to apply here as it is relatively easy to obtain a large amount of data to train the regression model.}. Repeating this process for each film $i$ allows us to compute its associated \textit{relative error}, $e_i$. A film's relative error is given by
\begin{displaymath}
e_i = \frac{|y_i - p_i|}{y_i},
\end{displaymath}
where $y_i$ is the actual amount film $i$ earned during its opening weekend, and $p_i$ is the prediction generated by the multivariate regression\footnote{We used the number of Wikipedia page views for each film seven days before its premier to create the relative error plots in the \textit{Results and Discussion} Section.}. A significant relative error indicates a discrepancy between a movie's predicted box office revenue and its actual revenue which is large \textit{relative} to the movie's actual box office revenue.

\section{Results and Discussion}
\label{rnd}
\subsection{United States Box Office}
\begin{figure}
\centering
\includegraphics[scale=0.45]{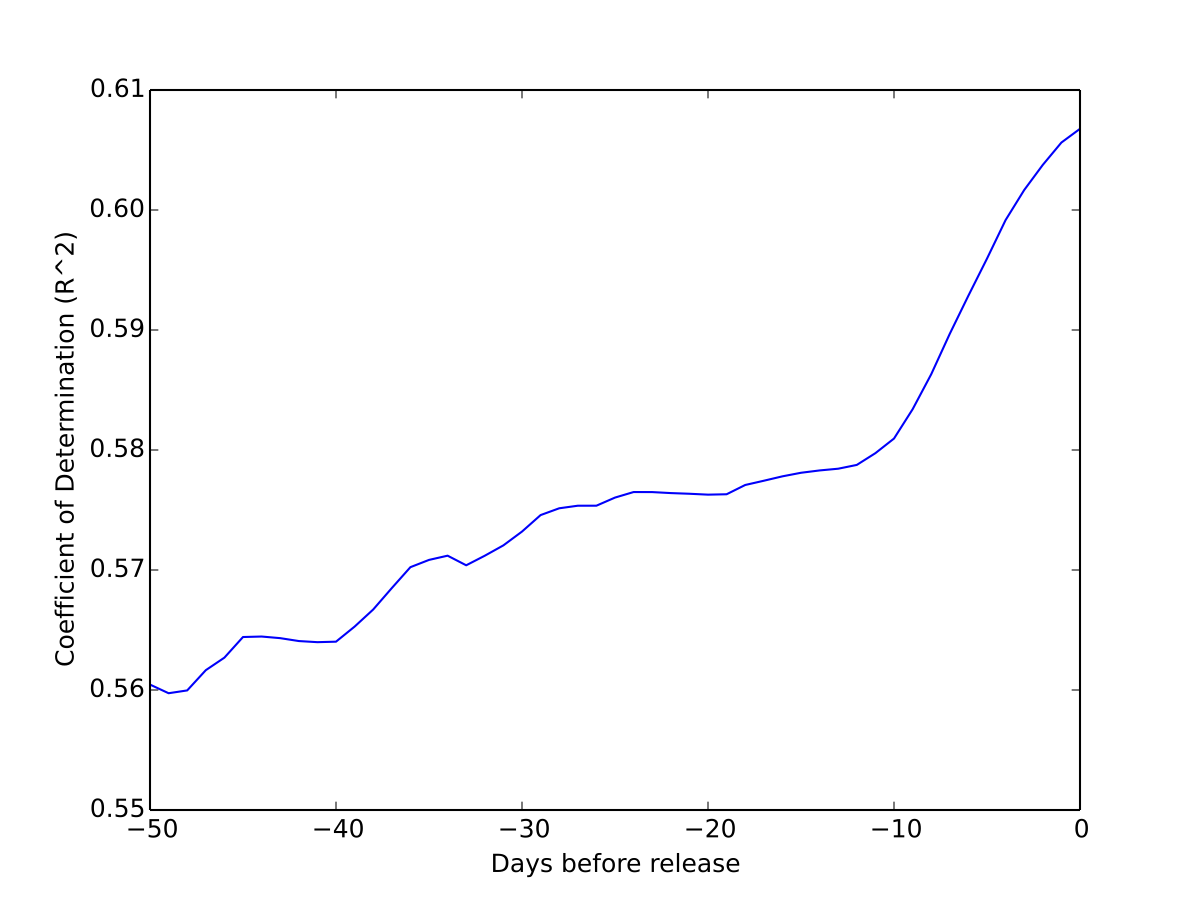}
\caption{Evolution of the R$^2$ coefficient of determination in time (US films)}
\label{us correlation}
\end{figure}

As was noted in the \textit{Methods} Section, once we had each US film's associated Wikipedia page view data along with its screen count, we formed a separate linear regression model on each day leading up to its premier.
Thus we were able to visualize the accuracy of our predictions as a function of time as in Figure \ref{us correlation}.

As one would expect, the accuracy of the model improves as the movie premiers draw near and movie awareness swells. This is the time period when potential moviegoers are presumably researching which movie to see during the coming weekend. Using this data set we obtain accuracy comparable to that reached by the model based only on Wikipedia page views used by Mesty{\'a}n et al.\footnote{Note that the coefficient of correlation plotted in \cite{mestyan:wikipedia} is the Pearson correlation, R, while we use R$^2$.}\cite{mestyan:wikipedia}. This is the level of accuracy we expected to obtain since the only major difference between the two models is that they were used on different lists of films.

\begin{figure}
\centering
\includegraphics[scale=0.2]{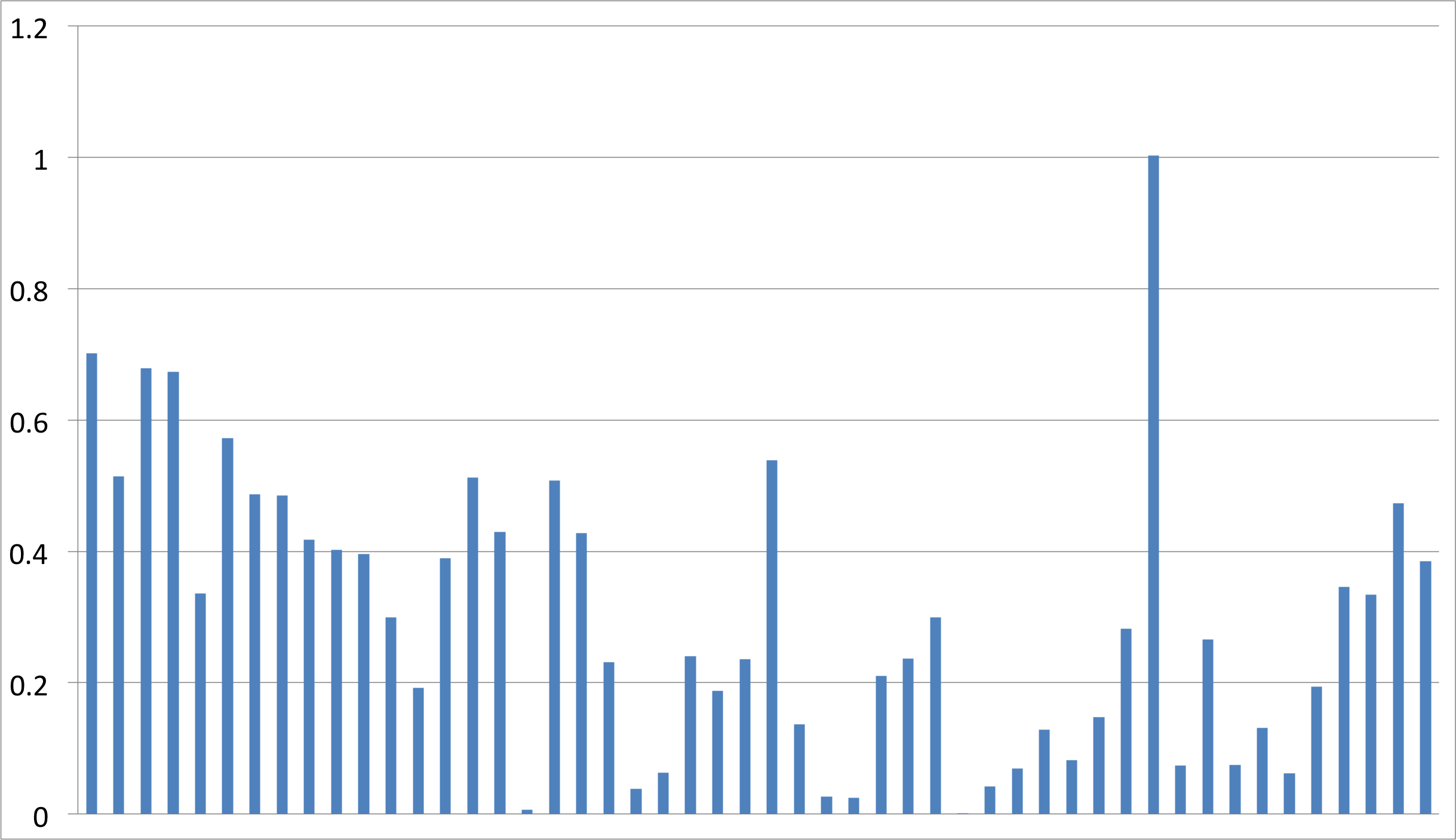}
\caption{Relative Error for 50 US films}
\label{us error}
\end{figure}

In Figure \ref{us error} we have the relative errors (see the \textit{Cross-Validation} Section) for the first 50 films from the US data set. In this case the films have been sorted in descending order based on their box office revenues. The relative errors for these films carry more weight than those of the movies that were left out. This is because, in computing the relative error, we divide the difference between the actual box office revenue and the prediction by the actual box office revenue. For example, over predicting by one million dollars the box office revenue for a film which brought in ten million dollars during its opening weekend would result in a relative error of $0.1$ while over predicting by the same amount the box office revenue for a film which only brought in one hundred thousand dollars would yield a much greater relative error of $9.0$. A large relative error for a film that did well during its opening weekend indicates a considerable gap between our prediction and its actual box office revenue. We see in Figure \ref{us error} that the relative errors are reasonably small, as we would expect considering the R$^2$ value achieved predicting United States box office results.

\subsection{Foreign Box Office}
After recreating one of the results of Mesty{\'a}n et al. we applied the same techniques to sets of films released in the United Kingdom, Australia, Germany, and Japan.

\subsubsection{English-Speaking Foreign Box Offices}
One possible problem with employing the same approach used to predict box office profits in the United States to anticipate those in the United Kingdom and Australia is that all three countries speak a common language. As a result internet users in all three nations use the same English branch of Wikipedia. This makes it impossible to discern which of the three countries a page view on a movie's Wikipedia page came from. Combined with the different release dates for films across the three markets, a large amount of noise is likely being introduced. For instance, if film ``A'' premiers in the United States before Australia, the United States premier could create an influx of Wikipedia page views earlier than normal, relative to Australia's premier. This, in turn, could cause the model to overestimate the film's expected box office revenue in the Australian market. Indeed there are over 90 films in both the lists from Australia and the United Kingdom which have different release dates from their counterparts in the United States.

\begin{figure}
\centering
\includegraphics[scale=0.45]{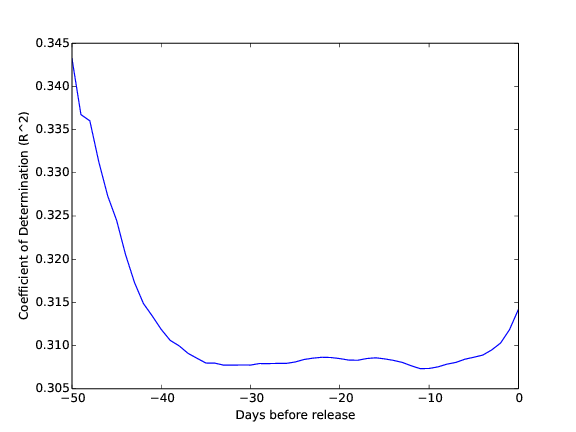}
\caption{Evolution of the R$^2$ coefficient of determination in time (UK films)}
\label{uk correlation}
\end{figure}

Regressing on the number of Wikipedia page views and screen counts for each of the $141$ UK films at each day leading up to the films' premiers, we produce a plot of the accuracy of our model as a function of time (Figure \ref{uk correlation}). The maximum value the R$^2$ coefficient of determination attains here is about $0.34$; much less than that for the United States predictions. The shape of the graph is also strikingly different from Figure \ref{us correlation}. The R$^2$ coefficient is at its maximum $50$ days before the film premiers and decreases as time progresses, eventually increasing marginally in the few days before the premiers. This strange behavior may be the result of differing movie release dates, as discussed above.

\begin{figure}
\centering
\includegraphics[scale=0.2]{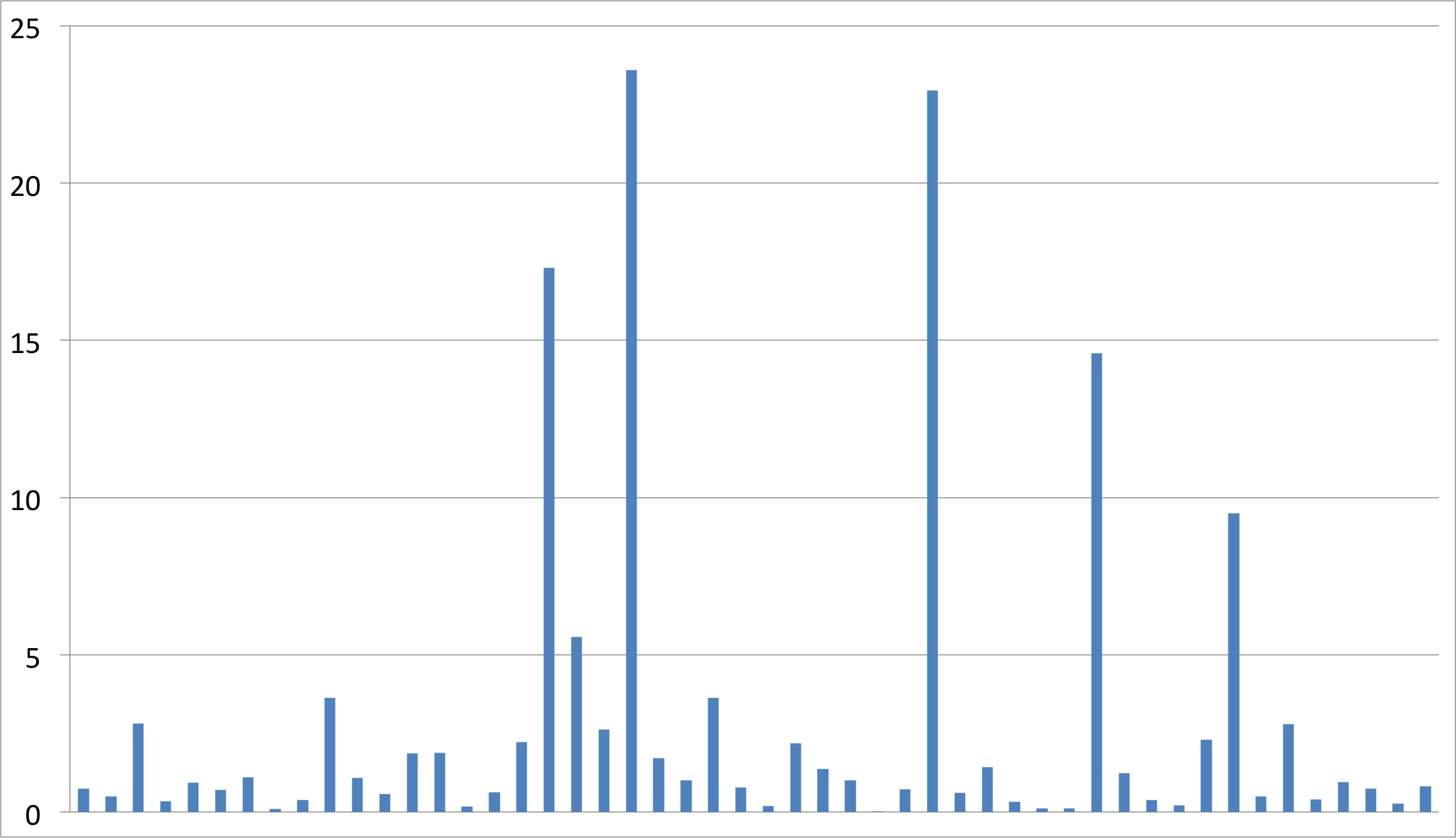}
\caption{Relative Error for 50 UK films}
\label{uk error}
\end{figure}

The relative errors for the films from the United Kingdom are given in Figure \ref{uk error}. There are a few outliers with very large relative errors, but the errors are larger on average, than for the US films. This is intuitive given that the coefficient of determination of the linear regression is much lower for the films from the United Kingdom.

\begin{figure}
\centering
\includegraphics[scale=0.45]{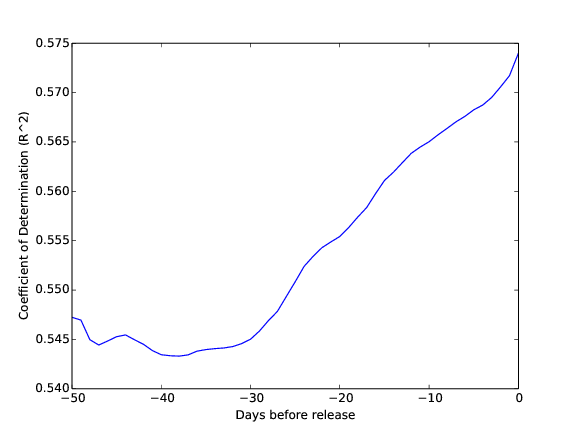}
\caption{Evolution of the R$^2$ coefficient of determination in time (Australian films)}
\label{au correlation}
\end{figure}

\begin{figure}
\centering
\includegraphics[scale=0.2]{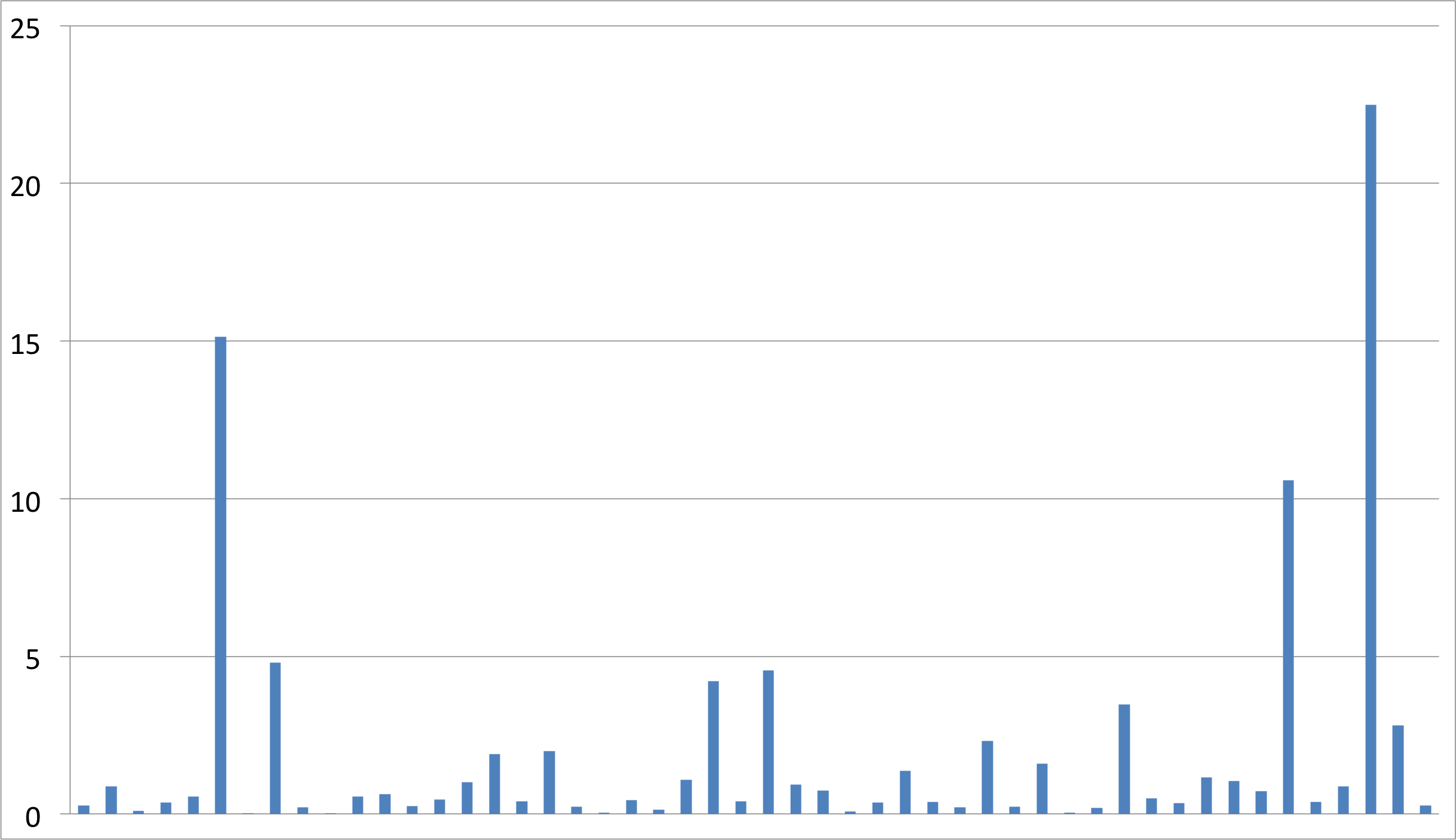}
\caption{Relative Error for 50 Australian films}
\label{au error}
\end{figure}

Repeating the steps taken for the films from the UK with the $118$ Australian movies, we obtain Figure \ref{au correlation}. Here the coefficient of determination evolves with time in a similar manner to that in Figure \ref{us correlation}; it increases as time does. The model attains a relatively high R$^2$ value of $0.57$ a day before the movie release dates. Figure \ref{au error} shows the relative error for 50 Australian films. There are fewer films with huge relative errors than we saw for the UK data set and the model produces lower relative errors, on average, for the Australian data set than the UK one.

\subsubsection{Japanese and German Speaking Box Offices}
Since Wikipedia has separate versions written entirely in Japanese and German, it is a reasonable assumption that each is accessed by primarily Japanese and German citizens, respectively. This resolves the stumbling block we faced with the English-speaking nations, namely determining which market each viewer belonged to. Hence one might expect that predictions for Japanese and German opening weekend box office revenue using Wikipedia page views would outperform those for the UK and Australia.

\begin{figure}
\centering
\includegraphics[scale=0.45]{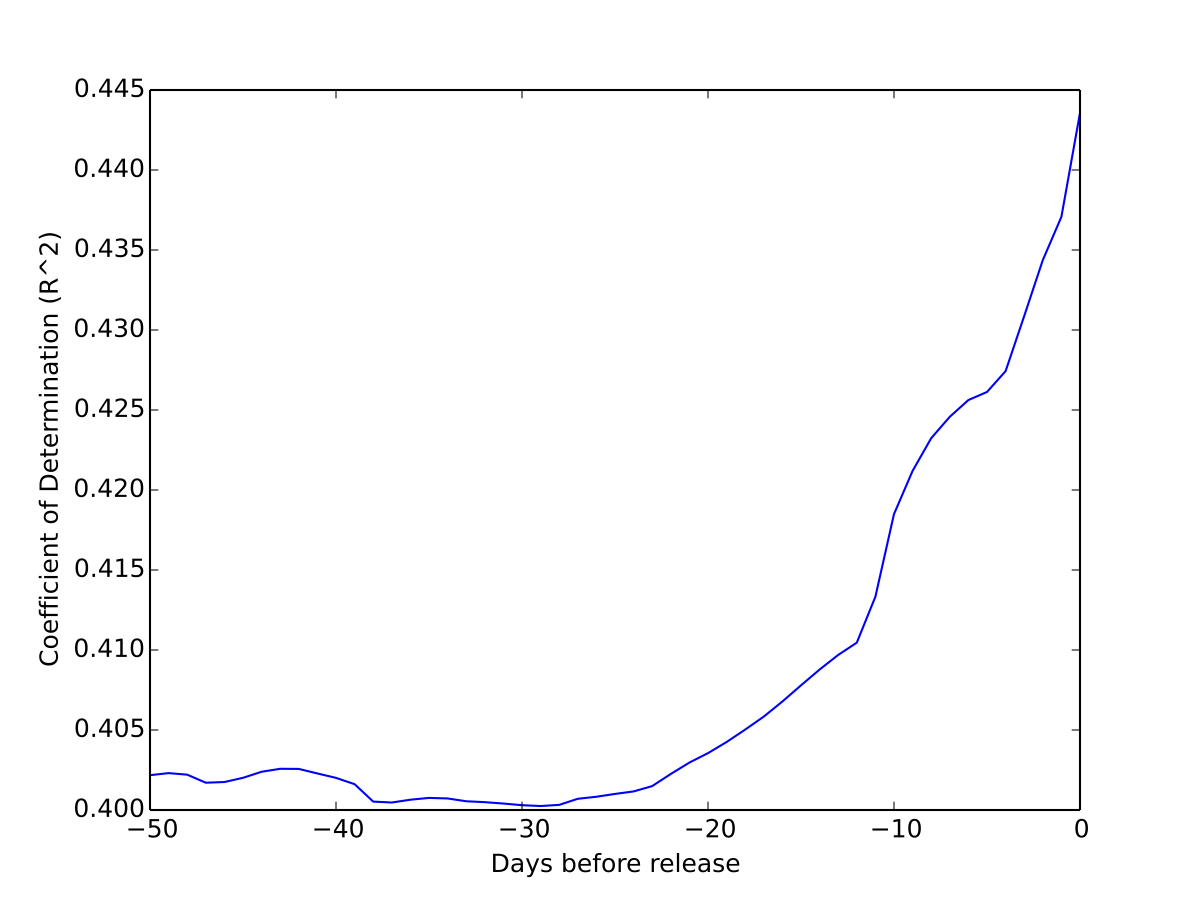}
\caption{Evolution of the R$^2$ coefficient of determination in time (German films)}
\label{de correlation}
\end{figure}

\begin{figure}
\centering
\includegraphics[scale=0.2]{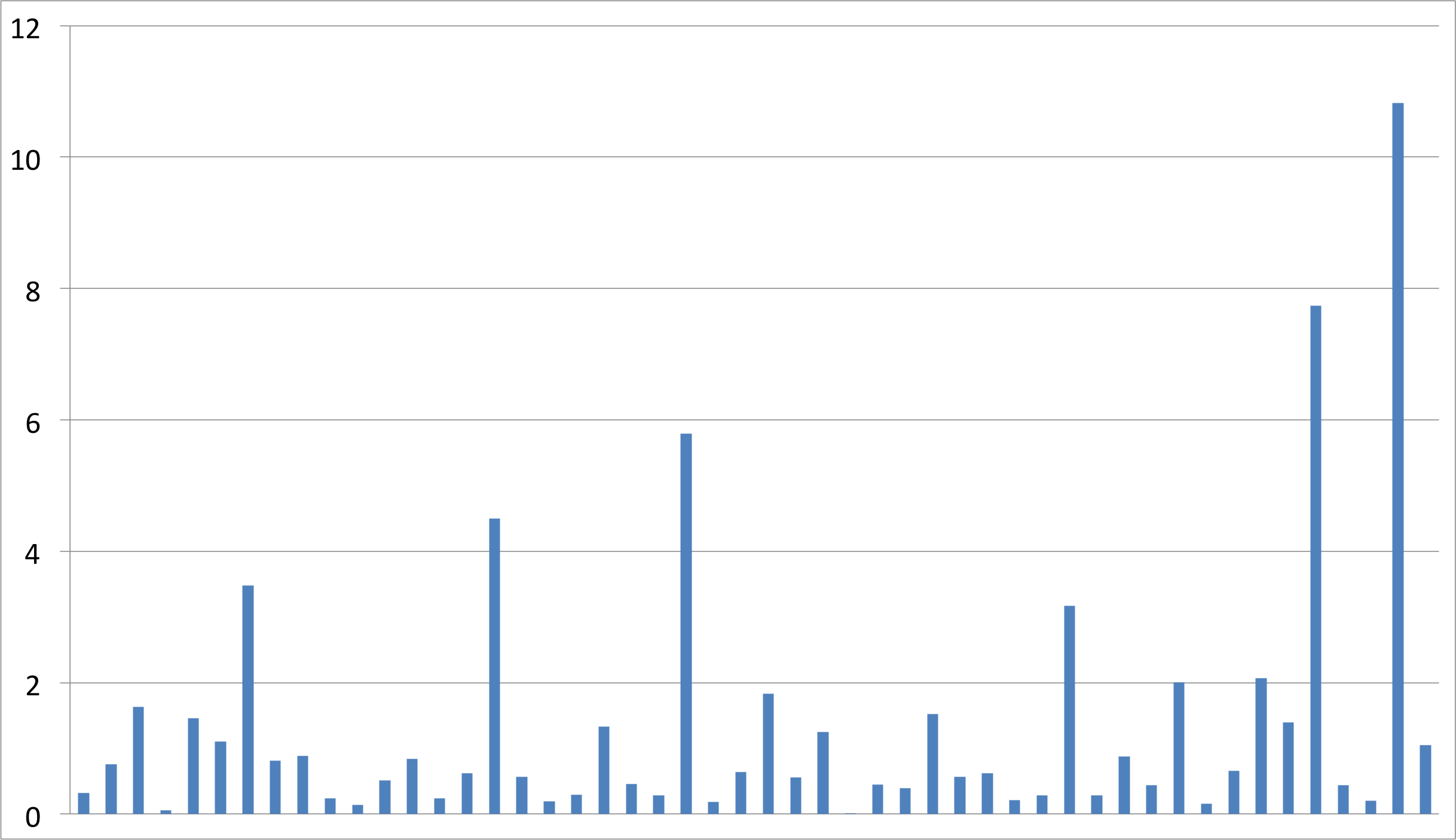}
\caption{Relative Error for 50 German films}
\label{de error}
\end{figure}

Carrying out the same procedure as before for the $105$ German films yields Figure \ref{de correlation}. While the correlation changes in a similar manner to that in Figure \ref{us correlation} (i.e. it increases as the release date approaches), the overall accuracy achieved is considerably lower. The maximum R$^2$ value attained is $0.45$, compared to the almost $0.61$ R$^2$ coefficient associated with the United States films. Figure \ref{de error} shows the relative error for the German films.

\begin{figure}
\centering
\includegraphics[scale=0.45]{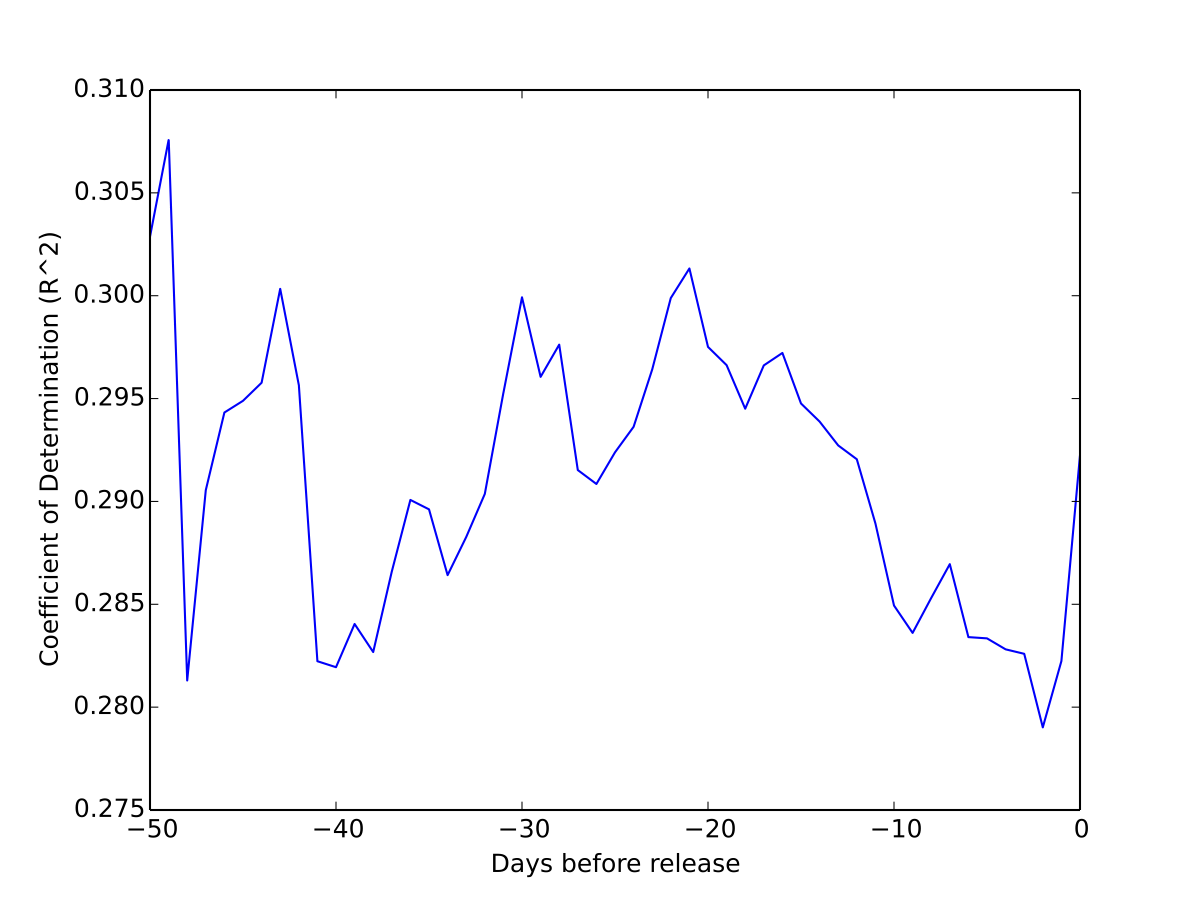}
\caption{Evolution of the R$^2$ coefficient of determination in time (Japanese films)}
\label{ja correlation}
\end{figure}

\begin{figure}
\centering
\includegraphics[scale=0.2]{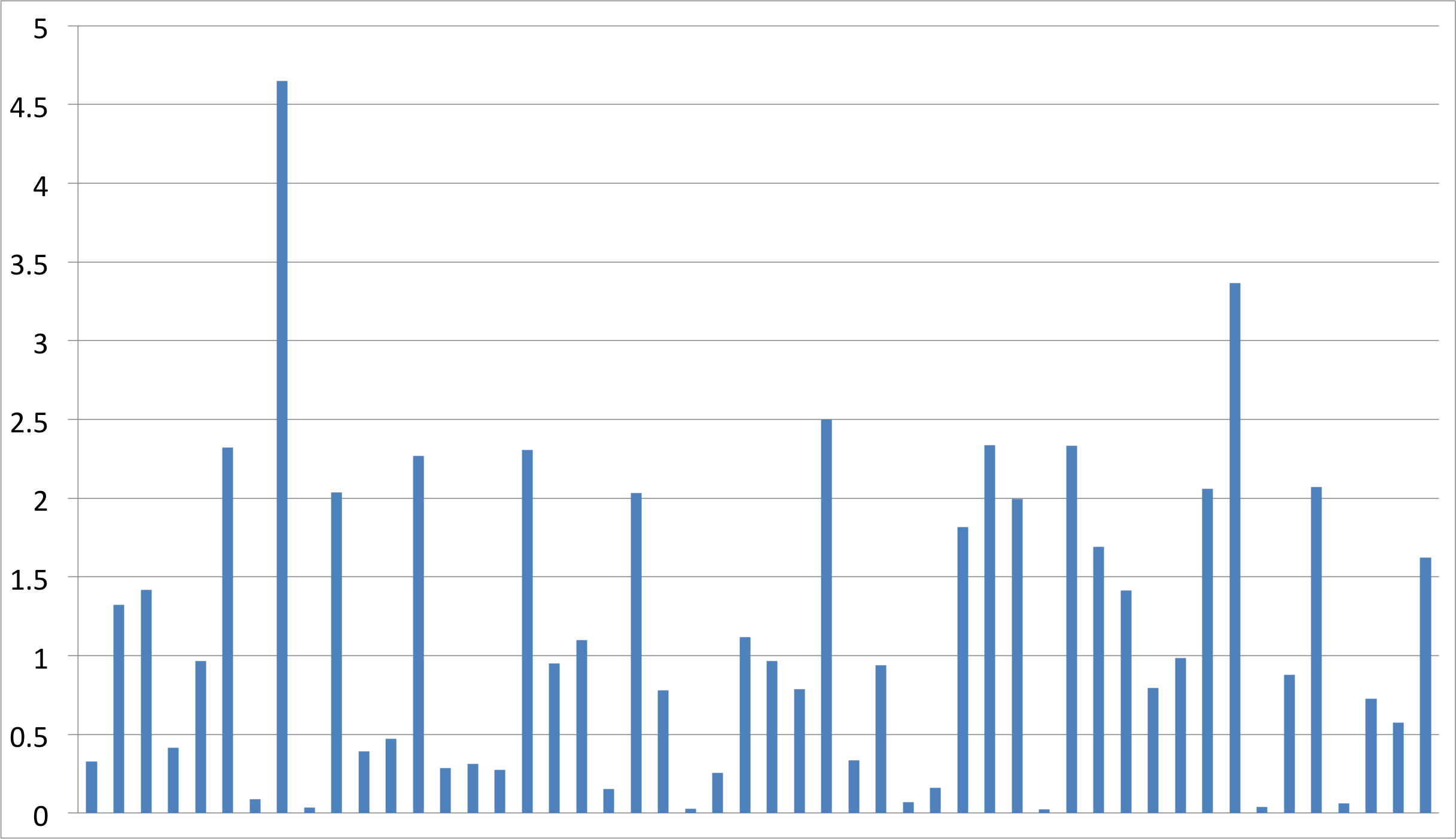}
\caption{Relative Error for 50 Japanese films}
\label{ja error}
\end{figure}

The results are even worse when the model is used to predict opening weekend box office revenues for $95$ films released in Japan. Figure \ref{ja correlation} shows the evolution of the coefficient of determination for this multivariate regression. The same pattern present in the United States', Australian, and German R$^2$ evolution plots does not appear here. The overall accuracy of the regression model is also markedly lower when it is applied to the Japanese market than the German market (maximum R$^2 \approx 0.30$). In Figure \ref{ja error} we have the relative error for the Japanese films. Using the regression model often predicts double the amount films actually earn during their opening weekends. This is unsurprising, given the low coefficient of determination attained with this set of films.\\

Overall, total Wikipedia page views do not appear to be strong predictors of how films will perform during their opening weekends in Germany, the UK, or Japan. It is likely that when researching which movie to see, citizens of Japan, Germany, and the UK simply do not use Wikipedia as a resource as often as people in the United States and Australia. Or, perhaps they use different means than the internet to aid their decisions altogether, or do not perform any research at all before going to the theaters. Another possible explanation for the disparity in accuracy between the predictions for the films released in the United States and those released in the UK, Germany, and Japan is the number of movies considered. Since there were two to three times as many films in the United States data set as the other sets, it included numerous low-budget, lesser-known movies. It is possible that the box office success of lesser-known movies is easier to predict. Indeed, if we remove the top 100 highest grossing films from the list of US films, our method obtains a maximum R$^2$ of almost $0.72$. However, this does not explain why we are able to predict box office revenue in Australia with more certainty than in other foreign countries. In any case, it is clear that a technique to predict opening weekend box office profits in the United Kingdom, Germany, or Japan must rely on more than Wikipedia page views alone.


\section{Conclusion}
In this article we have shown that although the method proposed by Mesty{\'a}n et al. predicts films' opening weekend box office revenues in the United States and Australia with reasonable accuracy, its performance drops significantly when applied to various foreign markets. In particular, we constructed a multivariate linear regression using the number of views the Wikipedia pages of various films released in the United States received to predict their successes during their opening weekends in theaters. We automated the process of determining a film's corresponding Wikipedia page so that we could apply the same technique to foreign movies. However, when we used the model to predict the opening weekend box office revenues generated by films in British, Japanese, and German theaters, we found its accuracy to be far from satisfactory. Finally we gave brief discussions of possible causes of the discrepancies.

While Wikipedia page views may be a strong predictor of box office performance for films in the United States, the same cannot necessarily be said for films released in other nations. Before a model similar to that presented here is used to predict box office sales in a foreign market, it should be tested on backdata gathered from that market.

\section{Acknowledgments}
The authors would like to thank Dean Shaw for his help interpreting Japanese film titles and URLs.

\balance

\bibliographystyle{acm-sigchi}
\bibliography{sigproc}
\end{document}